\documentclass[preprint2]{aastex}

\shorttitle{New chemo-kinematic relation}
\shortauthors{Minchev et al.}

\begin{document}

\title{A new stellar chemo-kinematic relation \\ reveals the merger history of the Milky Way disc}

\author{I.~Minchev\altaffilmark{1}, C.~Chiappini\altaffilmark{1}, M.~Martig\altaffilmark{2}, M. Steinmetz\altaffilmark{1}, R. S. de Jong\altaffilmark{1}, C. Boeche\altaffilmark{3}, C. Scannapieco\altaffilmark{1}, T. Zwitter\altaffilmark{4,5}, R. F. G. Wyse\altaffilmark{6}, J. J. Binney\altaffilmark{7}, J. Bland-Hawthorn\altaffilmark{8}, O. Bienaym\'{e}\altaffilmark{9}, B. Famaey\altaffilmark{9}, K.~C. Freeman\altaffilmark{10}, B. K. Gibson\altaffilmark{11}, E. K. Grebel\altaffilmark{3}, G. Gilmore\altaffilmark{12}, A. Helmi\altaffilmark{13}, G. Kordopatis\altaffilmark{12}, Y. S. Lee\altaffilmark{14}, U. Munari\altaffilmark{15}, J. F. Navarro\altaffilmark{16}, Q. A. Parker\altaffilmark{17,18,19}, A. C. Quillen\altaffilmark{20}, W. A. Reid\altaffilmark{17,18}, A. Siebert\altaffilmark{9}, A. Siviero\altaffilmark{1,21}, G. Seabroke\altaffilmark{22}, F. Watson\altaffilmark{19}, M. Williams\altaffilmark{1}
}

\altaffiltext{1}{Leibniz-Institut f\"{u}r Astrophysik Potsdam (AIP), An der Sternwarte 16, D-14482, Potsdam, Germany}
\altaffiltext{2}{Centre for Astrophysics \& Supercomputing, Swinburne University of Technology, P.O. Box 218, Hawthorn, VIC 3122, Australia}
\altaffiltext{3}{Astronomisches Rechen-Institut, Zentrum f\"{u}r Astronomie der Universit\"{a}t Heidelberg, M\"{o}nchhofstr. 12-14, D-69120 Heidelberg, Germany}
\altaffiltext{4}{University of Ljubljana, Faculty of Mathematics and Physics, Jadranska 19, 1000 Ljubljana, Slovenia}
\altaffiltext{5}{Center of Excellence SPACE-SI, Askerceva 12, 1000 Ljubljana, Slovenia}
\altaffiltext{6}{Department of Physics and Astronomy, Johns Hopkins University, 3400 North Charles Street, Baltimore, MD 21218, USA}
\altaffiltext{7}{Rudolf Peierls Centre for Theoretical Physics, Keble Road, Oxford OX1 3NP, UK}
\altaffiltext{8}{Sydney Institute for Astronomy, School of Physics, University of Sydney, NSW 2006, Australia}
\altaffiltext{9}{Universit\'e de Strasbourg, CNRS, Observatoire Astronomique, 11 rue de l'Universit\'e, 67000 Strasbourg, France}
\altaffiltext{10}{Australian National University, Canberra, Australia}
\altaffiltext{11}{Jeremiah Horrocks Institute, University of Central Lancashire, Preston, PR1 2HE, UK}
\altaffiltext{12}{Institute of Astronomy, University of Cambridge, Madingley Road, Cambridge CB3 0HA, UK}
\altaffiltext{13}{Kapteyn Astronomical Institute, University of Groningen, P.O. Box 800, 9700 AV Groningen, The Netherlands}
\altaffiltext{14}{Department of Astronomy, New Mexico State University, Las Cruces, NM 88003, USA}
\altaffiltext{15}{INAF National Institute of Astrophysics, Astronomical Institute of Padova, 36012 Asiago (VI), Italy}
\altaffiltext{16}{Senior CIfAR Fellow, University of Victoria, P.O. Box 3055, Station CSC, Victoria, BC V8W 3P6, Canada}
\altaffiltext{17}{Department of Physics \& Astronomy, Macquarie University, Sydney, NSW 2109 Australia}
\altaffiltext{18}{Research Centre for Astronomy, Astrophysics and Astrophotonics, Macquarie University, Sydney, NSW 2109, Australia}
\altaffiltext{19}{Australian Astronomical Observatory, PO Box 915, North Ryde, NSW 1670, Australia}
\altaffiltext{20}{(Department of Physics and Astronomy, University of Rochester), Rochester, NY 14627}
\altaffiltext{21}{Department of Physics and Astronomy, Padova University, Vicolo dell Osservatorio 2, 35122 Padova, Italy}
\altaffiltext{22}{Mullard Space Science Laboratory, University College London, Holmbury St Mary, Dorking, RH5 6NT, UK}

\begin{abstract}
The velocity dispersions of stars near the Sun are known to increase with stellar age, but age can be difficult to determine so a proxy like the abundance of $\alpha$ elements (e.g., Mg) with respect to iron, [$\alpha$/Fe], is used. Here we report an unexpected behavior found in the velocity dispersion of a sample of giant stars from the RAdial Velocity Experiment (RAVE) survey with high quality chemical and kinematical information, in that it decreases strongly for stars with [Mg/Fe]~$>0.4$~dex (i.e., those that formed in the first Gyr of the Galaxy's life). These findings can be explained by perturbations from massive mergers in the early Universe, which have affected more strongly the outer parts of the disc, and the subsequent radial migration of stars with cooler kinematics from the inner disc. Similar reversed trends in velocity dispersion are also found for different metallicity subpopulations. Our results suggest that the Milky Way disc merger history can be recovered by relating the observed chemo-kinematic relations to the properties of past merger events.

\end{abstract}

\keywords{Galaxy: disk --- Galaxy: evolution --- Galaxy: formation --- Galaxy: abundances  --- Galaxy: kinematics and dynamics --- solar neighborhood}

\section{Introduction}

\begin{figure*}
\epsscale{2.10}
\plotone{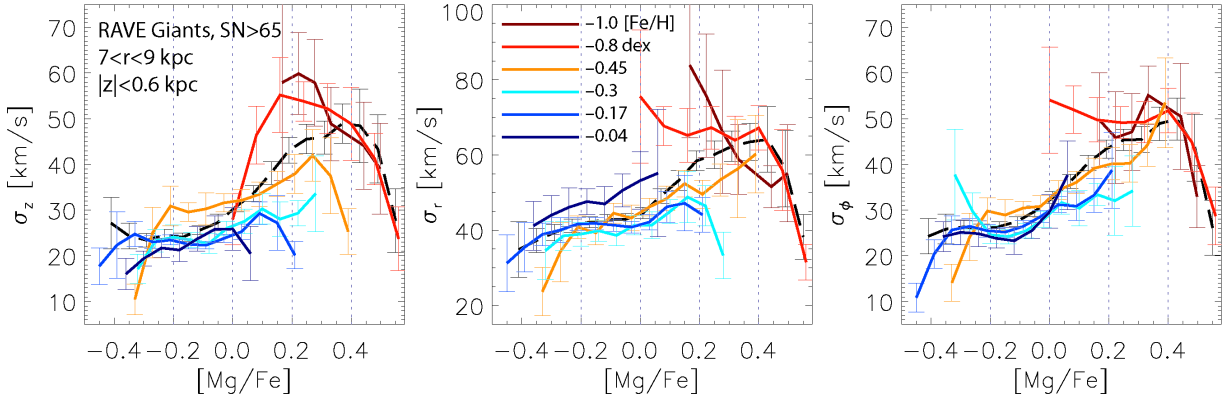}
\caption{
{\bf Left:} Vertical velocity dispersion $\sigma_z$, as a function of [Mg/Fe] ratios for RAVE giants. The black dashed curve shows the total sample. The color-coded curves present subpopulations grouped by common median metallicity as indicated in the middle panel, with resulting mean values [Fe/H]~$= -1.05, -0.85, -0.45, -0.3, -0.1, +0.125$~dex. The error for each [Mg/Fe] bin is estimated as the two standard deviations of 1000 realizations in a bootstrapping calculation. The number of stars in the three highest-[Mg/Fe] bins are 43, 25, and 11, respectively. Because variation in both chemistry and kinematics is expected with change in position in the Galactic disc, we constrain our sample to Galactocentric distances in the range $7<r<9$~kpc and consider a maximum vertical height above and below the disc plane $|z|=0.6$~kpc, where r and z are the radial and vertical coordinates in a cylindrical system. {\bf Middle:} Same as on the left, but for the radial velocity dispersion $\sigma_r$. {\bf Right:} Same as on the left and middle, but for the azimuthal velocity dispersion $\sigma_\phi$. Similar reversal in the velocity dispersion trends at [Mg/Fe]~$ > 0.4$~dex is found for all velocity components.
}
\label{fig:fig1}
\end{figure*}

Understanding galaxy formation and evolution is one of the central goals of contemporary astronomy and cosmology. High-redshift observations provide insight into the evolution of global galaxy properties, but are fundamentally limited in probing the internal kinematics and chemistry on sub-galactic scales. The Milky Way is the only galaxy within which we can obtain information at the level of detail required to understand these processes. This realization is manifested in the number of on-going and planned spectroscopic Milky Way surveys, such as RAVE \citep{steinmetz06}, SEGUE \citep{yanny09}, APOGEE \citep{Majewski10}, HERMES \citep{freeman10}, Gaia-ESO \citep{gilmore12}, Gaia \citep{perryman01}, and 4MOST \citep{dejong12}, which aim at obtaining kinematic and chemical information for a large number of stars. Despite the increasing amounts of observational data, to date we have lacked the means to discriminate among different thick-disc formation scenarios and to understand the disc merger history in general. Here we fill this gap.

Stars near the Sun have long been known to follow an age-velocity relation, where velocity dispersions increase with age \citep{wielen77}. Due to the lack of good age estimates, the shape of the age-velocity relation has been a matter of debate \citep{freeman91, binney00, seabroke07}. Alternatively, the [$\alpha$/Fe] ratios, such as [Mg/Fe] can be used to identify the oldest stars because the interstellar medium is quickly enriched in elements formed by $\alpha$-capture in short-lived massive stars while there is a delay before iron-peak elements are produced in abundance by thermonuclear supernovae (SNIa) \citep{matteucci12, haywood13}. 

Studying the SEGUE G-dwarf sample, \cite{bovy12} and \cite{liu12} have recently argued that the stellar vertical velocity dispersion increases for populations of decreasing metallicity and increasing [$\alpha$/Fe], but this relation has been somewhat unclear at the low-metallicity end. We re-examine the $\sigma_z$-[Fe/H]-[$\alpha$/Fe] connection by presenting it in a different way compared to earlier works: instead of color maps, here we overlay curves of velocity dispersion with [Mg/Fe] for only six different metallicity bins allowing for better statistics in the low metallicity regime where the number of stars gets small.

\section{Results}

\subsection{Chemo-kinematic relation in RAVE}

We study data from the RAdial Velocity Experiment (RAVE) \citep{steinmetz06, zwitter08, siebert11a, kordopatis13}. RAVE is a magnitude-limited survey of stars randomly selected in the $9 < I < 12$ magnitude range and is presently the largest spectroscopic sample of stars in the Milky Way for which individual elemental abundances are available \citep{boeche11}.

For the present work we selected a RAVE sample of giant stars, which is close to a random magnitude-limited sample. We excluded giants with log g~$< 0.5$ to avoid any possible effects due to the boundaries of the learning grid used for the automated parameterization and considered stars in the temperature range $\rm 4000<T_{eff} < 5500$ (thus avoiding horizontal branch stars, see \citealt{boeche13}).

The estimated total measurement uncertainties (internal plus external) of the new RAVE-DR4 pipeline for giants (log g~$< 3.5$~dex) with SN~$> 50$ in the temperature and metallicity range we consider in the present work ($-1.5 <$~[Fe/H]~$ < 0.5$) are (see Table 2 in \citealt{kordopatis13}) $\sim100$~K in $\rm T_{eff}$, $\sim0.3$~dex for log g, and $\sim0.1$ for the metallicity [M/H]. The final uncertainties in [Fe/H] and [Mg/Fe] computed by the chemical pipeline are $\sim0.1$ and 0.15~dex, respectively. To further decrease these, we select a sample of giants with SN~$>65$, resulting in 4755 stars with high-quality chemistry and kinematics. The mean uncertainties in galactocentric velocities are 10-15 km/s.

In the left panel of Fig.~\ref{fig:fig1} the black-dashed curve shows the variation with [Mg/Fe] of the vertical velocity dispersion $\sigma_z$. Contrary to the expected monotone increase of $\sigma_z$ with $\alpha$-enhancement, we see a strong inversion in this trend at [Mg/Fe]~$>0.4$~dex. This is unexpected because stars with higher [$\alpha$/Fe] ratios, at a given radius, should be older and, thus, should possess larger random energies. The six metallicity ([Fe/H]) bins, shown by color-coded curves, allow us to probe deeper. We see that for [Fe/H] $> -0.4$, $\sigma_z$ is independent of [Fe/H] and a slowly increasing function of [Mg/Fe] but that for more metal-poor stars $\sigma_z$ increases with decreasing [Fe/H]. An interesting observation is the distinct decrease in $\sigma_z$ at the high-[Mg/Fe] end of all but one [Fe/H] subpopulation, with the $\sigma_z$-maximum shifting to lower [Mg/Fe] for bins of increasing metallicity. For all velocity components, the most metal-poor, $\alpha$-rich stars (which must be the oldest) have velocity dispersions comparable to those of the most metal-rich (youngest) populations. In Sec.~\ref{sec:sim} we show that this puzzling behavior can be explained by radial migration of old cold stellar samples from the inner disc.

\begin{figure*}
\epsscale{2.1}
\plotone{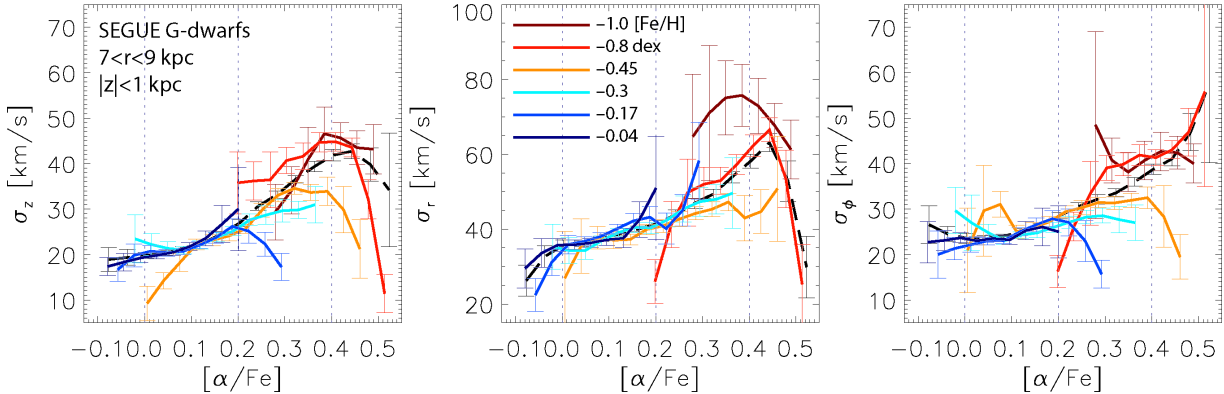}
\caption{
\label{fig:fig2}
Similar to Fig.~\ref{fig:fig1}, but for the SEGUE G-dwarf sample. Same median metallicity are used with resulting mean values [Fe/H]~$= -1.0, - 0.81, -0.45, -0.28, -0.1, +0.06$~dex (very close to RAVE). While we constrain our sample to the same Galactocentric distance range as in Fig.~\ref{fig:fig1} and \ref{fig:fig3} ($7 < r < 9$~kpc), here we consider a maximum vertical height above and below the disc plane $|z| = 1$~kpc instead of 0.6~kpc, because of the lack of stars with $|z| < 0.5$~kpc in SEGUE.
}
\end{figure*}

\subsection{Comparison to SEGUE G-dwarfs}

The results we obtained here with RAVE data should be seen in other ongoing/near-future Galactic surveys. We check this by studying the currently available SEGUE G-dwarf sample described by \cite{lee11}, by considering stars with SN~$>30$. Due to the lack of individual chemical element measurements in SEGUE, we use their [$\alpha$/Fe], which is the average of several $\alpha$-elements. Because of the low number of stars at $|z| < 0.5$~kpc in SEGUE, a maximum distance above and below the disc plane of $|z| = 1$~kpc is used (instead of $|z| = 0.6$~kpc). This results in a sample comprising $\sim$~10300 stars. Fig.~\ref{fig:fig2} presents the same information as Fig.~\ref{fig:fig1}, but for SEGUE. We recover the expected strong decline in the vertical and radial velocity dispersions, $\sigma_z$ and $\sigma_r$, for the most metal-poor samples. While no reversal is found in the dispersion of the azimuthal velocity component, $\sigma_\phi$, for the total sample, the expected trend is seen for the intermediate metallicity sub-populations (blue and orange curves) and hinted in the most metal-poor bin (maroon). This assures that our findings are not pertinent to RAVE data only, but should be a general result. 

We note that the same trend is hinted in Fig.2 by \cite{bovy12}, but less clear possibly due to the cut of SN~$>15$, larger [$\alpha$/Fe] bins, and larger disk coverage more prone to erase structure that may vary spatially.

\begin{figure*}
\epsscale{2.10}
\plotone{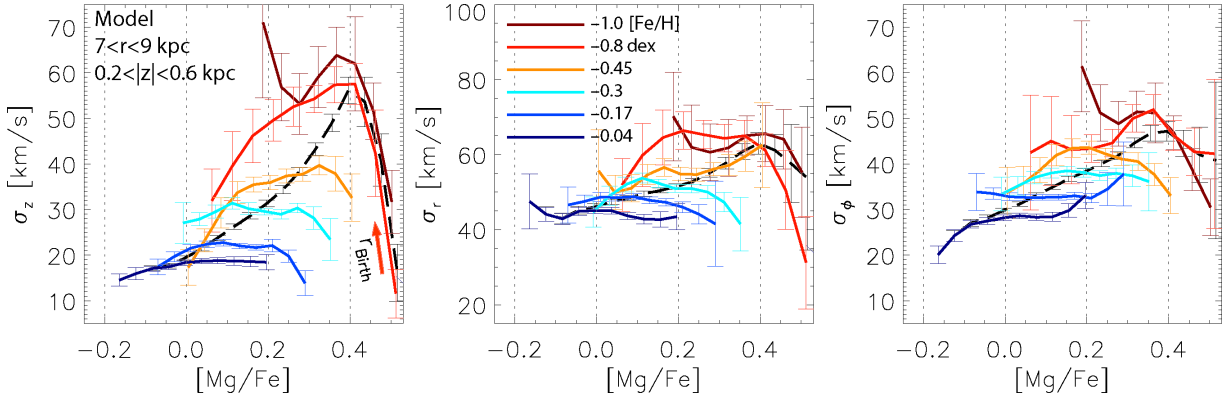}
\caption{
Same as Fig.~\ref{fig:fig1}, but for a chemo-dynamical model incorporating early massive mergers. Uncertainties of $\delta$[Fe/H]~$=\pm0.12$~dex and $\delta$[Mg/Fe]~$=\pm0.14$~dex are convolved into the simulated data. We use the same radial range as in RAVE and constrain vertically the sample within $0.2 < |z| < 0.6$~kpc, since RAVE misses the stars closest to the disc plane. The median metallicity bins are the same as in the data and the resulting mean values are [Fe/H]~$= -1.1, -0.82, -0.44, -0.27, -0.1, +0.07$~dex (very close to the data). Similar trends are seen as in the data, where samples in most metallicity bins decrease their velocity dispersions at the high-[Mg/Fe] end. 
\label{fig:fig3}
}
\end{figure*}

\section{Interpretation}
\label{sec:sim}

The highest velocity dispersions ($\sigma_z, \sigma_\phi> 40$~km/s and $\sigma_r > 60$~km/s) achieved by the lowest  three metallicity bins are too large to be accounted for by internal disc evolution processes, such as scattering by giant molecular clouds \citep{lacey84}, the spiral arms \citep{jenkins92}, or the Galactic bar \citep{minchev12a,minchev12b}. On the other hand, perturbations by mergers at the early times of disc formation \citep{wyse01,villalobos08,quinn93} clumpy disc instabilities \citep{bournaud09}, stars born hot at high redshift \citep{brook05,brook12}, and/or accreted stellar populations \citep{abadi03,meza05} can be a plausible explanation. 

We test the possibility that mergers of decreasing satellite-to-disc-mass ratios and frequency (including stars born hot at high redshift) have caused the trends we find in RAVE giant stars, by using simulated data from a chemo-dynamical model \citep{mcm13} built by the fusion between a high-resolution galaxy simulation in the cosmological context with Milky Way characteristics and a detailed chemical evolution model. A strong perturbation from the last massive merger (1:5 disc mass ratio 8-9 Gyr before the present) plays an important role in the disc evolution. Due to its in-plane orbit (inclination less than $45^\circ$), this event drives strong radial migration by triggering spiral structure. Similar, but less intense events occur throughout the disc evolution. 

Fig.~\ref{fig:fig3} presents the same information as Figures~\ref{fig:fig1} and \ref{fig:fig2}, but for the model. A very good agreement is found between the data and model for the velocity dispersion trends of different metallicity subpopulations.

The stars with the highest [$\alpha$/Fe]-ratios and lowest [Fe/H] form in the inner disc at the onset of disc formation in an inside-out disc formation scenario. Radial migration can then bring these to the solar vicinity, contaminating the locally evolved sample. Because of the very fast initial chemical evolution, stars with [Mg/Fe] $> 0.25$~dex and [Fe/H]~$< -0.5$~dex have approximately the same age, as shown in Fig.~\ref{fig:fig4}, top. These stars, being the oldest disc population, have the longest time available for migration to the solar vicinity. However, because migration efficiency decreases with increasing velocity dispersion, stars with cold kinematics would be affected the most \citep{sellwood02}. This is the case for the stars in the metallicity bin with median [Fe/H]~$= -0.8$~dex (red curve), which are seen to migrate the largest distances (Fig.~\ref{fig:fig4}, bottom).

\subsection{Radial migration cools the disc during mergers}

\begin{figure}
\epsscale{.8}
\plotone{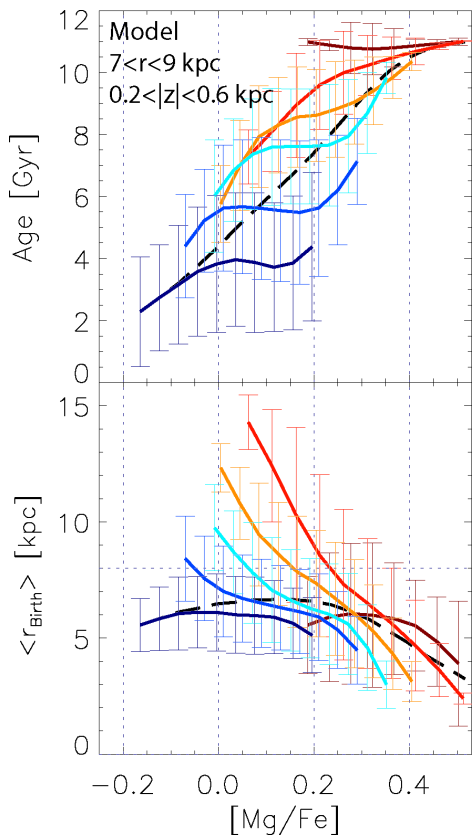}
\caption{
Origin of stars currently in the solar neighborhood. {\bf Top:} Mean age as a function of [Mg/Fe] in the simulated solar neighborhood for the same metallicity bins used for the data and model. An almost linear relation between age and [Mg/Fe] is found up to [Mg/Fe]~$=0.4$~dex. However, this is not the case for a given sub-sample, where the mean age is mostly constant and decreasing/increasing at the low/high-[Mg/Fe] ends, respectively. Error bars of two standard deviations show the scatter around the mean. {\bf Bottom:} Mean birth-radius, $\langle r_{\rm Birth}\rangle$, as a function of [Mg/Fe]. The dotted horizontal line at 8~kpc indicates the solar radius. The spread in birth radius decreases as more metal-rich (younger) populations are considered, as expected. The inversion in the velocity dispersion trends at [Mg/Fe]~$>0.4$~dex results from stars born at mean $\langle r_{\rm Birth}\rangle < 5-6$~kpc. Note that the velocity dispersion decline at the high-[Mg/Fe] end of each metallicity sub-sample is also caused by stars with older mean ages born at progressively smaller radii.
\label{fig:fig4}
}
\end{figure}

Figures~\ref{fig:fig3} and \ref{fig:fig4} showed that the decline in velocity dispersion for stars with the highest [$\alpha$/Fe] ratios in RAVE and SEGUE data can be interpreted as the effect of stars born in the innermost disc. For this to occur, some of the oldest disc stars must have migrated from the inner disc to the solar neighborhood, while remaining on near-circular orbits, significantly cooler than coeval samples born at progressively larger radii.

\begin{figure}
\epsscale{1.}
\plotone{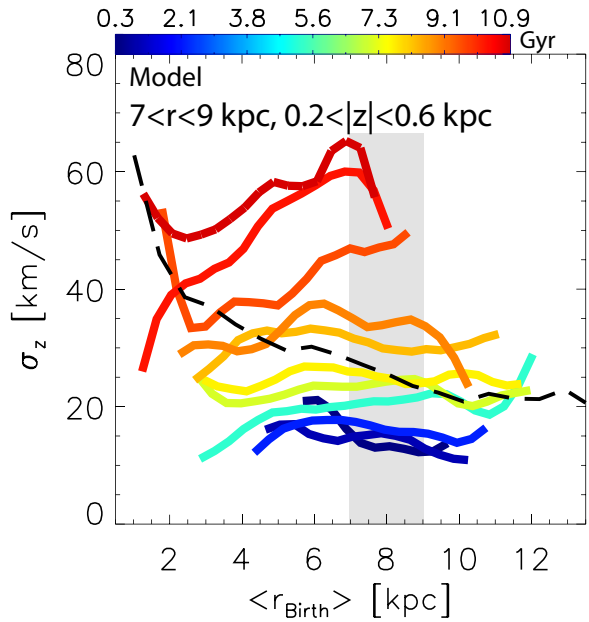}
\caption{
\label{fig:fig5}
Vertical velocity dispersion, $\sigma_z$, as a function of mean birth radius, $\langle r_{\rm Birth}\rangle$, for the simulated solar neighborhood sample used in the previous figures. The black-dashed line shows the total population. Different colours correspond to different age groups. The decreasing range in $\langle r_{\rm Birth}\rangle$ with age is related to the decreasing time available for radial migration. For the oldest samples stars arrive at the solar vicinity much cooler than stars born in-situ, due to the stronger effect of mergers on the outer disc and the decreasing probability of migration with increasing velocity dispersion. The coolest/oldest samples arriving from the innermost disc are the equivalent to the most metal-poor, [$\alpha$/Fe]-rich stars in RAVE, SEGUE and the chemo-dynamical model. The large positive gradient found for old populations (red colors) turns into a negative slope for the youngest samples (dark blue), indicating a quiescent regime, where stars arriving from the inner disc heat slightly the local velocity distribution.
}
\end{figure}

In Fig.~\ref{fig:fig5} we now check whether this is indeed that case, using the same simulated sample as in all previous figures. The black-dashed curve shows that, for the total sample, the final vertical velocity dispersion, $\sigma_z$, increases with decreasing mean birth radius, $\langle r_{\rm Birth}\rangle$. Naively, this could be interpreted as evidence that the local disc heats as a result of the migration. However, by decomposing this sample into mono-age populations, we find that the continuous increase in $\sigma_z$ with decreasing $\langle r_{\rm Birth}\rangle$ is caused by the growing fraction of younger stars born near their current locations, i.e., the solar circle. The fact that stars migrating from the inner disc are generally hotter is related to their older ages allowing them to be exposed to perturbations causing both heating \citep{wielen77} and migration \citep{sellwood02}. However, for most coeval populations (color curves) there exists the general trend of $\sigma_z$ decreasing with decreasing $\langle r_{\rm Birth}\rangle$, which becomes more important for older samples. For the next-to-oldest population, stars originating near the galactic center arrive to the solar vicinity colder by ~35 km/s compared to those born near the Sun. It is clearly seen that stars arriving from the inner disc can be both cooler and older than a locally born sample, which explains the observations.

Fig.~\ref{fig:fig5} illustrated that that cooler/warmer populations can arrive from the inner/outer disc during mergers but this trend reverses as external perturbations become unimportant. Therefore, there must exist a critical time at which the disc enters from the first regime into the second. Fig.~\ref{fig:fig5} shows that in our model this reversal occurs at  $\sim$age~$<1.5$~Gyr (dark blue lines). Future work can try to relate this to observations using chemical information.

\subsection{Implications for the Milky Way disc merger history}

From the above discussion it follows that the decrease in the velocity dispersion for the highest-[Mg/Fe] stars in the model is a sign that they have arrived in the solar vicinity kinematically cooler than the locally born coeval population. This conclusion is remarkable, as stars migrating outward due to internal instabilities alone are expected to heat the disc (\citealt{loebman11,schonrich09b}, even if only weakly -- \citealt{minchev12b}). The opposite behavior we find here must be related to the presence of mergers in our simulation. Indeed, Fig.~\ref{fig:fig4} shows that metal-poor stars with [Mg/Fe]~$> 0.4$~dex (which cause the decline in velocity dispersion for the model - red curves in Fig.~\ref{fig:fig3}) arrive from progressively smaller galactic radii, while having approximately the same age. Similarly to the most metal-poor stars, the inversion in the $\sigma_z$-[Mg/Fe] relation for the other [Fe/H] subpopulation can be related to mergers of diminishing strength perturbing the Milky Way disc throughout its lifetime. Within this interpretation, the properties of these events can be recovered in a differential study. For example, satellite-to-disc-mass ratios and orbital parameters can be related to the mean velocity dispersions of metallicity bins and the inversion in the velocity dispersion profiles as functions of [Mg/Fe], for each velocity component.

\section{Discussion}

In this work we presented a new chemo-kinematic relation in RAVE giants, where a decline in the velocity dispersion was found for stars with [Mg/Fe]~$>0.4$~dex, as well as for samples in narrow ranges of [Fe/H]. We verified that another big data set -- the SEGUE G-dwarfs -- shows similar relations.  
By comparing to a chemodynamical model, we explained these results as the stronger effect of mergers on the outer parts of discs and the subsequent radial migration of older stellar populations with cooler kinematics born in the inner disc. 

The interpretation of our results -- that we see the effect of satellite perturbations on the disc -- is not strongly model dependent because (i) mergers are expected to always affect more significantly the outer disc \citep{bournaud09}, (ii) a general prediction of inside-out-formation chemical evolution models is a fast decrease in [$\alpha$/Fe] with increasing radius for stars in a narrow metallicity range, at high redshift \citep{matteucci12}, and (iii) radial migration has been firmly accepted to be inseparable part of disc evolution in numerical simulations \citep{sellwood02,mf10,roskar12}. It may be possible that debris accreted from a disrupted galaxy \citep{abadi03,meza05}, which should possess the highest velocity dispersions in the disc, had just the right chemistry to create the maximum we observe at [Mg/Fe] = 0.4 dex, although to explain the $\sigma_z$-maxima of all [Fe/H] bins this interpretation requires a series of mergers of exactly the right increasing metallicity. Furthermore, accretion would not explain the exceedingly low velocity dispersion values for the most $\alpha$-rich stars, which must be linked to radial migration of stars born in the inner disc. The gas-rich turbulent clumpy-disc formation scenario for the thick disc may also provide a viable explanation but in conjunction with perturbations from mergers (making it similar to our model and \citealt{brook12}), necessary to produce the observed decline in velocity dispersion at the high-[$\alpha$/Fe] end of each metallicity subpopulation. Finally, a quiescent evolution scenario (e.g., \citealt{schonrich09b}), where the Milky Way thick disc formed by heating an initially thin disc through internal evolution processes only, is unfeasible, given that in that case stars arriving from the inner disc should be slightly hotter than the locally born population \citep{loebman11,minchev12b}, i.e., the contrary to what the observations suggest.

The stars with the lowest [Fe/H] and highest [Mg/Fe] ratios identified in this work possess the chemistry and kinematics, which allow them to be associated with the oldest Milky Way population born in the bar/bulge region. Although currently efforts are made to look for these in the inner Galaxy, we have shown that, thanks to radial migration, one can also study them here in the solar vicinity.

While the Milky Way has been seen as an unusually quiet galaxy in view of the predictions by the $\rm \Lambda$CDM theory \citep{hammer07}, the results of this work suggests that, in addition to being important for the formation of the thick disc (e.g., \citealt{wyse01, mcm13}), satellite-disc encounters of decreasing intensity were at play throughout its evolution.

To secure the conclusions of this work, follow-up high-resolution spectroscopic observations are needed for the high [Mg/Fe], low velocity dispersion populations.

With the availability of stellar ages in the near future from the Gaia mission, it will become possible to confirm or rule out the currently proposed explanation for the observed chemo-kinematic relations, by requiring that the data are also reproduced by Fig.~\ref{fig:fig4}. If confirmed in the number of ongoing and forthcoming Galactic surveys, our discovery would provide a missing piece in the current understanding of the Milky Way disc evolution.

\acknowledgments
Funding for RAVE has been provided by: the Australian Astronomical Observatory; the Leibniz-Institut f\"{u}r Astrophysik Potsdam (AIP); the Australian National University; the Australian Research Council; the French National Research Agency; the German Research Foundation (SPP 1177 and SFB 881); the European Research Council (ERC-StG 240271 Galactica); the Istituto Nazionale di Astrofisica at Padova; The Johns Hopkins University; the National Science Foundation of the USA (AST-0908326); the W. M. Keck foundation; the Macquarie University; the Netherlands Research School for Astronomy; the Natural Sciences and Engineering Research Council of Canada; the Slovenian Research Agency; the Swiss National Science Foundation; the Science \& Technology Facilities Council of the UK; Opticon; Strasbourg Observatory; and the Universities of Groningen, Heidelberg and Sydney. The RAVE web site is at http://www.rave-survey.org.

\end{document}